# Transit timing analysis of the exoplanet TrES-5 b. Possible existence of the exoplanet TrES-5 c

Eugene N. Sokov[1,2], Iraida A. Sokova[2], Vladimir V. Dyachenko[1], Denis A. Rastegaev[1], Artem Burdanov[3], Sergey A. Rusov[2], Paul Benni[4], Stan Shadick[5], Veli-Pekka Hentunen[6], Mark Salisbury[7], Nicolas Esseiva[8], Joe Garlitz[9], Marc Bretton[10], Yenal Ogmen[11] , Yuri Karavaev[12], Anthony Ayiomamitis[13], Oleg Mazurenko[14], David Molina Alonso[15], Velichko Sergey[16,17]

[1] *Special Astrophysical Observatory, Russian Academy of Sciences, Nizhnij Arkhyz, Russia, 369167;*
[2] *Central Astronomical Observatory at Pulkovo of Russian Academy of Sciences, Pulkovskoje shosse d. 65, St. Petersburg, Russia, 196140*;
[3] *Space sciences, Technologies and Astrophysics Research (STAR) Institute, Université de Liège, Allée du 6 Août 17, 4000 Liège, Belgium*
[4] *Acton Sky Portal (Private Observatory), Acton, MA, USA;*
[5] *Physics and Engineering Physics Department, University of Saskatchewan, Saskatoon, SK, S7N 5E2, Canada;*
[6] *Taurus Hill Observatory, Warkauden Kassiopeia ry., Härkämäentie 88, 79480 Kangaslampi, Finland;*
[7] *School of Physical Sciences, The Open University, Milton Keynes, MK7 6AA, UK;*
[8] *Observatory Saint Martin, code k27, Amathay Vésigneux, France;*
[9] *Private Observatory, 1155, Hartford St, Elgin, Oregon 97827, USA;*
[10] *Baronnies Provençales Observatory, Hautes Alpes - Parc Naturel Régional des Baronnies Provençales, 05150 Moydans, France;*
[11] *Green Island Observatory, Code B34, Gecitkale, Famagusta, North Cyprus;*
[12] *Institute of Solar-Terrestrial Physics (ISTP), Russian Academy of Sciences, SiberianBranch, Irkutsk, Russia;*
[13] *Perseus Observatory, Athens, Greece;*
[14] *Trottier Observatory, Physics Department, SFU, Burnaby, BC, Canada;*
[15] *Anunaki Observatory, AstroHenares Association, Rivas Vaciamadrid, Madrid, Spain;*
[16] *Institute of Astronomy, Kharkov V.N. Karazin National University, Kharkov, Ukraine;*
[17] *International Center for Astronomical, Medical and Ecological Research NAS of Ukraine, Kyiv, Ukraine*

**Abstract.** In this work, we present transit timing variations detected for the exoplanet TrES-5b. To obtain the necessary amount of photometric data for this exoplanet, we have organized an international campaign to search for exoplanets based on the Transit Timing Variation method (TTV) and as a result of this we collected 30 new light curves, 15 light curves from the Exoplanet Transit Database (ETD) and 8 light curves from the literature for the timing analysis of the exoplanet TrES-5b. We have detected timing variations with a semi-amplitude of A ≈ 0.0016 days and a period of P ≈ 99 days. We carried out the N-body modeling based on the three-body problem. The detected perturbation of TrES-5b may be caused by a second exoplanet in the TrES-5 system. We have calculated the possible mass and resonance of the object: $M \approx 0.24M_{Jup}$ at a 1:2 Resonance.

## 1 Introduction

There are many methods of searching for exoplanets. The radial velocity and transit photometry methods are the main ones, because most of exoplanet discoveries were made using these two methods (based on statistics from exoplanets.org and exoplanets.eu (Schneider et al. 2011; Han et al. 2014). These techniques most often lead to the discovery of the closest exoplanets, such as hot Jupiters and Saturn type exoplanets around solar-type stars, due to the more apparent interaction of the planet with its host star, which is easily detected in just a short period of time.

Despite this, new exoplanets on more distant orbits in known exoplanet systems are being discovered every year. One of the methods which allows us to predict or discover other exoplanets in known discovered planetary systems is the Transit Timing Variation method (TTVs) described by Miralda-Escude 2002; Agol et. al. 2005; Narita 2009 and Hoyer 2011. This method is based on the periodic variation of the planet's orbit around the parent star manifesting itself as a delay or advance of the moment of the middle of transit, caused by the gravitational influence of another planet or some other more massive object also orbiting around the star.

The first exoplanet with well-detected timing was Kepler-19b. With a period of about 300 days and a semi-amplitude equal to 5 minutes, the exoplanet Kepler-19c was predicted (Ballard S., et al., 2011). Following this, the existence of Kepler 19c was confirmed by the radial velocities method (Malavolta et al., 2017). Further exoplanets Kepler-46c (Nesvorny et. al, 2012), Kepler-419c (Rebekah & Dawson et. al., 2014), Kepler-338e (Eylen & Albrecht, 2015), KOI-620.02 (Masuda, 2014) have been also discovered by the TTV method.

In recent years, with the increase of the quality and quantity of photometric observations during the exoplanet transits, the search for extrasolar planets by means of the Transit Timing Variations method (TTVs) has become more effective and relevant.

In this work, we describe the organization of the international observational campaign and the investigation of the detected timing variations of TrES-5b. The exoplanet TrES-5b orbits a cool G dwarf GSC 03949-00967 (V = 13.72 mag) and was discovered by the Trans-Atlantic Exoplanet Survey in 2011 (Mandushev, et. al., 2011). The orbital period of the exoplanet TrES-5b predicted in that work is $P$ = 1.4822446 +/- 0.0000007 days. The mass and radius of TrES-5b are $M_{pl}$ = 1.778 (± 0.063) $M_{Jup}$, $R_{pl}$ = 1.209 (± 0.021) $R_{Jup}$ (Mandushev et. al., 2011).

Earlier, the exoplanet TrES-5b was investigated by (Mislis et al., 2015) and (Maciejewski et al., 2016). In these papers, the information on the orbital parameters of TrES-5b, such as the orbital period $P$, orbital inclination $i_b$, radius of planet in stellar radii $R_b/R_*$, semi-major axis in stellar radii $a_b/R_*$ have been refined. At the same time, although evidence of timing (TTV) for TrES-5b has not been detected, there is also no convincing evidence of its absence.

## 2 The speckle interferometry observations

In November of 2015 and June of 2016, high precision imaging of the star TrES-5 was carried out with the 6-meter BTA telescope (Special Astrophysical Observatory) using a speckle interferometer. We used an EMCCD (electron-multiplying CCD) to take images with the BTA speckle interferometer. Thus, an image of a faint object represents a set of separate points where the light quanta fall.

The main contribution to the optical image distortion and blurring belongs to the atmospheric turbulence (or atmospheric seeing). For example, for a 6-m aperture of the optical BTA telescope at the wavelength of 550 nm, the diffraction limit of resolution for a point source must be equal to 0.02″, whereas the real size of the image influenced by the atmospheric effects amounts to 1–2″, i.e. 100 times more. The speckle interferometry method is a method of observing astronomical objects seen through a turbulent atmosphere with the angular resolution limit close to the diffraction limit.

The principle of the speckle interferometry method is to take high-resolution images with a very short exposure time (~$10^{-2}$ s). Such images consist of a great number of speckles that are produced by the mutual interference of the light beams that fall on the focal plane of a telescope from different parts of the lens. Each speckle looks like an airy disk in the focal plane of a perfect telescope that is not affected by the atmospheric seeing. Atmospheric seeing influences the image in such a way that a wavefront that reaches a ground-based telescope is always distorted by the optical imperfections of the atmosphere. When taking very short-exposure images we record the speckle distribution at that very instant, while with long exposures the image loses its structure and becomes blurred. In the images of a non-point (extended) source, the speckle pattern (their shape and size) reflects the characteristics of the source itself. For example, if we observe a binary object (a binary star or a binary asteroid), then the speckles are recorded in pairs, and each pair of speckles represents an airy disk from the two components of a binary star or asteroid. In order to obtain information about the structure of the observed object we accumulated thousands of its snapshots.

Based on two observational sets of speckle interferometry of TrES-5b, two autocorrelation functions of the speckle-interferometry images were obtained. Because the star is faint (V=13.7 mag), the signal-to-noise ratio of the obtained measurements is low precision. Nevertheless, based on the results of two sets of TrES-5 observations, we can argue that there are no components near the star with a brightness difference of about Δm: 0mag ÷ 1mag and at a distance in the range of ρ: 200 mas ÷ 3000 mas, which corresponds to the range: 72 AU ÷ 1080 AU. Both autocorrelation functions are presented in Fig 1.

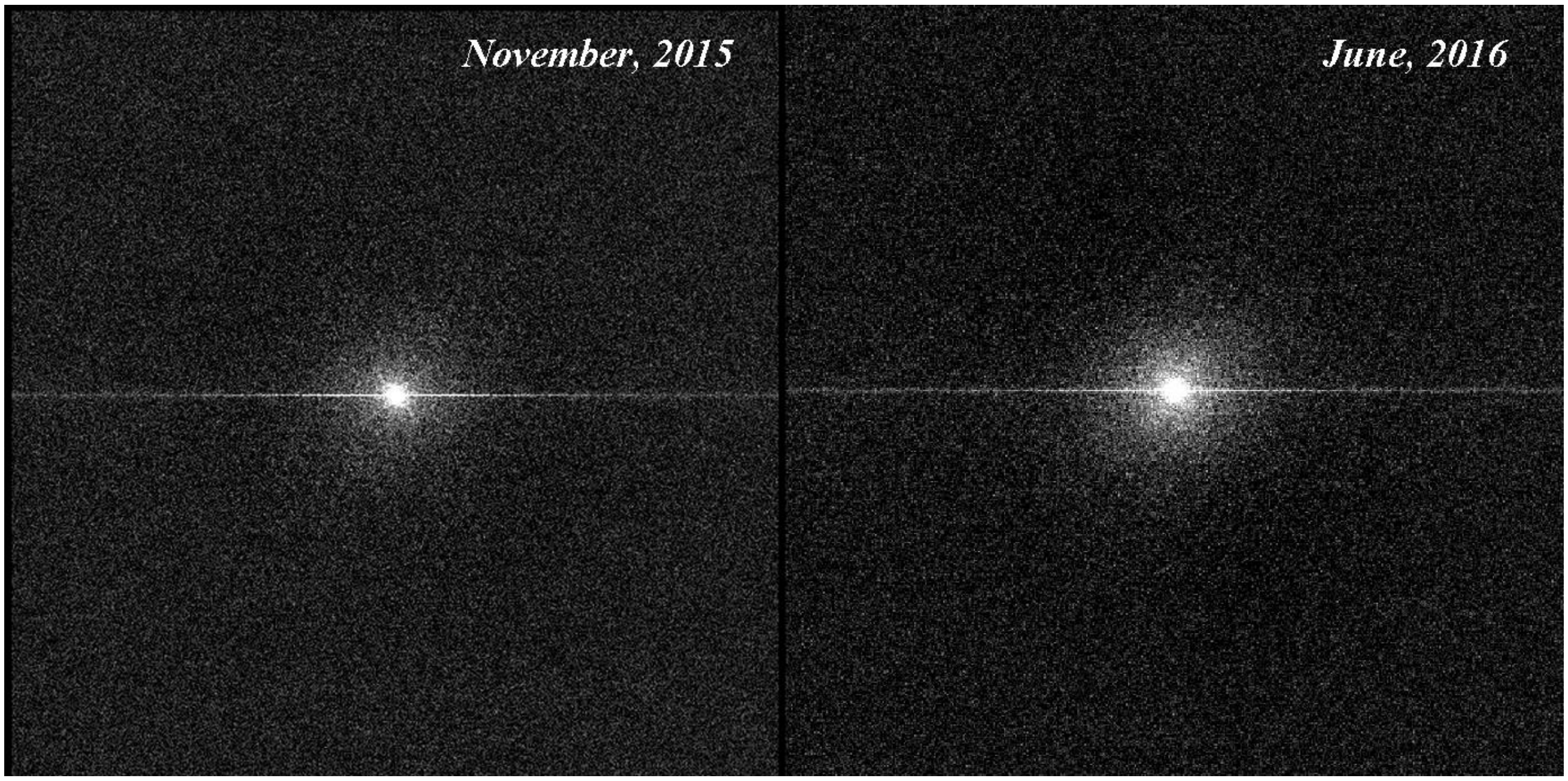

Fig 1. The autocorrelation function of speckle-interferometric images of TrES-5 (obtained on November, 2015 and June, 2016 with the use of 6-m BTA telescope)

## 3 Photometric observations

For the exoplanet search by the TTV method an international observation campaign as part of EXPANSION (**EX**o**P**lanetary tr**AN**sit **S**earch with an **I**nternation **O**bservational **N**etwork) project was organized. Observatories from Russia, Europe, North and South America with a small and middle diameter of telescopes from 25-cm to 2-m were used for the photometric observations of TrES-5b transits. All the telescopes participated in observational campaign are presented in Table 1.

Table 1. Telescopes participating in the observational campaign.

| Telescope | Aperture | Location |
|---|---|---|
| MTM-500M | 0.5m | Pulkovo Observatory (Kislovodsk), Russia |
| ZA-320M | 0.32m | Pulkovo Observatory (Saint-Petersburg), Russia |
| Zeiss-600 | 0.6m | ISTP SB RAS, Mondy, Russia |
| Ritchey-Chretien system | 0.82m | Baronnies Provencales Observatory, France |
| Cassegrain system | 0.43m | |
| Zeiss-2000 | 2.0m | IC AMER, Peak Terskol, Russia |
| Meade 14" LX200R | 0.35m | Famagusta, Cyprus |
| Meade 16" ACF OTA | 0.406m | Varkaus, Finland |
| Celestron C14 OTA | 0.36m | |
| Celestron C11EdgeHD | 0.28m | Amathay Vésigneux, France |
| Celestron C11EdgeHD | 0.28m | Acton, MA USA |
| Newton system | 0.3m | Elgin, OR USA |
| Optimised Dall Kirkham system | 0.4m | London, Great Britain |
| PlaneWave CDK700 | 0.7m | Trottier Observatory, Burnaby, Canada |
| Meade 8" LX200GPSR | 0.203m | Madrid, Spain |

Based on the campaign data we obtained 30 new light curves of the transits of TrES-5b. Due to the fact that the host star is quite faint for small and medium aperture telescopes, the star was observed predominantly without the use of filters to increase the SNR. In some cases, $R_c$ and $V_c$ filters of the Johnson-Cousins photometric system were used. The observation log is presented in Table 2.

Table 2. Details on new observations reported in this paper.

| Date (UT) | Telescope, aperture | Filter | *X* (airmass change) | Cadence, min |
|---|---|---|---|---|
| 2013-09-23 | CelestronC11, 0.28m | None | 1.04→1.29 | 0.47 |
| 2013-10-07 | Newton system, 0.3m | None | 1.03→1.50 | 0.98 |
| 2013-11-02 | Optimised Dall Kirkham system, 0.4m | $R_c$ | 1.13→1.47 | 0.45 |
| 2014-02-25 | ZA-320M, 0.32m | None | 1.72→1.22 | 0.54 |
| 2014-04-17 | MTM-500M, 0.5m | None | 2.05→1.31 | 1.1 |
| 2014-09-30 | MTM-500M, 0.5m | None | 1.47→2.1 | 1.12 |
| 2014-10-03 | MTM-500M, 0.5m | None | 1.58→1.7 | 1.5 |
| 2014-10-03 | Zeiss-2000, 2m | None | 1.57→1.7 | 2.0 |
| 2015-02-15 | Celestron C14 OTA,0.36m | None | 1.57→1.93 | 0.46 |
| 2015-03-09 | MTM-500M, 0.5m | None | 1.12→1.44 | 0.64 |
| 2015-05-18 | Zeiss-600, 0.6m | $R_c$ | 1.65→1.22 | 0.87 |
| 2015-05-29 | Ritchey-Chretien system, 0.82m | V | 1.29→1.04 | 0.47 |
| 2015-06-15 | Meade 14" LX200R, 0.35m | None | 1.95→1.2 | 0.83 |
| 2015-06-18 | Meade 14" LX200R, 0.35m | None | 1.97→1.3 | 0.84 |
| 2015-07-19 | Ritchey-Chretien system, 0.82m | None | 1.18→1.04 | 0.32 |
| 2015-08-31 | Meade 16ACF OTA, 0.406m | None | 1.0→1.18 | 0.54 |
| 2016-03-19 | Meade 16ACF OTA, 0.406m | None | 1.46→1.12 | 0.46 |
| 2016-03-21 | Meade 16ACF OTA, 0.406m | None | 1.61→1.23 | 0.46 |
| 2016-03-27 | Meade 16ACF OTA, 0.406m | None | 1.82→1.3 | 0.43 |
| 2016-04-02 | Meade 16ACF OTA, 0.406m | None | 1.9→1.51 | 0.45 |
| 2016-06-10 | Ritchey-Chretien system, 0.82m | None | 1.12→1.05 | 0.48 |
| 2016-06-26 | PlaneWave CDK700, 0.7m | None | 1.17→1.02 | 0.49 |
| 2016-07-25 | Meade 8" LX200GPSR, 0.203m | None | 1.21→1.11 | 0.24 |
| 2016-07-28 | Cassegrain system, 0.43m | None | 1.12→1.03 | 0.36 |
| 2016-07-31 | Meade 14" LX200R, 0.35m | None | 1.26→1.16 | 0.81 |
| 2016-08-14 | PlaneWave CDK700, 0.7m | None | 1.08→1.25 | 0.93 |
| 2016-10-28 | Celestron C11EdgeHD, 0.28m | None | 1.03→1.35 | 0.28 |
| 2016-10-31 | Ritchey-Chretien system, 0.82m | None | 1.05→1.22 | 0.45 |
| 2016-12-13 | Ritchey-Chretien system, 0.82m | None | 1.19→1.81 | 0.44 |
| 2017-05-16 | Ritchey-Chretien system, 0.82m | $I_c$ | 1.89→1.30 | 0.45 |

## 3 Data reduction and analysis of light curves

The photometric observations obtained in the campaign were processed by APEX-II, MuniWin, or AIP4Win software. The APEX-II package (Devyatkin et al., 2009) is completely automatic and has many options for the processing of astrometric and photometric observations. This package allows the use of aperture photometry and PSF photometry (PSF fitting). The MuniWin (Hroch, 2014) as well as AIP4Win (Tsamis et al., 2013) packages provide easy-to-use tools for all astronomical astrometry and photometry as well as FITS files operations and a simple user interface along with a powerful processing engine.

For all observational sets bias, flat-field and dark calibration images were obtained and subsequently taken into account in photometric data processing. When processing the photometric observations, for each series we chose, as a rule, 5–10 reference stars with brightness close to that of the object located on the frame close to it to reduce the effect of the atmospheric extinction. Based on the processing results, we studied the behavior of each reference star. If one of them was variable or if its behavior differed sharply from that of all the remaining stars, then it was excluded from the subsequent processing. The mean values between the derived magnitudes of the reference stars and the magnitude of the object were the sought-for a result — the object's light curve. The precision of the observations was determined using a check star that was chosen from the reference stars and was closest in brightness to the object. We performed the same procedure with the check star as that with the object — we found the mean difference between its brightness and the brightness of the remaining reference stars and calculated the standard deviation for the derived light curve, which was considered to be the accuracy of the observations. Thus, we plotted the final light curve obtained based on carried out differential photometry of stars with the smallest standard deviation.

After data processing, we obtained 30 light curves of the TrES-5b transits. We also obtained 15 light curves selected from the Exoplanet Transit Database (http://var2.astro.cz/ETD/) We have taken the light curves with DQ (Data Quality) ≤ 3 based on ETD standards, showing only a full transit, as well as clearly defined moments of transit ingress and egress. We have not considered partial transits because the midpoint of a transit may be determined incorrectly due to a possible presence of small but appreciable deviations of the transit durations. Thus, the total 45 light curves of transits of the exoplanet Tres-5b were prepared for further fitting and analysis. All light curves were detrended against the airmass changes. Time scales of all data series have been converted into the Barycentric Julian Date (BJD) format.

Each transit light curve was modelled with the online EXOFAST applet (Eastman et al., 2013) available on the NASA Exoplanet Archive (https://exoplanetarchive.ipac.caltech.edu/index.html). The Exoplanet Archive's version of EXOFAST offers IDL-based calculations as the original code of EXOFAST, and also provides sufficient back-end computing resources to enable Markov Chain Monte Carlo (MCMC) analysis. The fitting and analysis of light curves in the best-fitting model allow one to get a time of the mid of transit $T_{mid}$, radius of planet to stellar radii $R_b/R_*$ ratio, LD coefficients *u1* and *u2* of the quadratic law, orbital inclination $i_b$ and total duration of a transit $T_{Dur}$.

In order to calculate the limb darkening (LD) coefficient in EXOFAST, a band had to be selected. In those cases where observations were carried out without filters, the average wavelength in the sensitivity curve of the CCD camera was determined. Thus, the closest band of sensitivity of photometric observations for each telescope was determined.

The following initial parameters were used for the light curve fitting: surface gravity for assumed mass log $g = 4.517 \pm 0.012$, effective temperature $T_{eff} = 5171 \pm 36$, metallicity [Fe/H] = $0.2 \pm 0.1$ and the prior detected orbital period of TrES-5b $P_b = 1.4822446 \pm 0.0000007$ days (Mandushev et. al., 2011). The final 30 light curves obtained in the observational campaign with the superimposed model curves after the fitting and the residuals from the best-fit model are presented in Fig. 2(a) and Fig. 2(b).

We re-determined the orbital period $P_b$ = 1.482247063 ± 0.0000005 days. For the determination of *O-C* (Observation – Calculation) value we calculated the difference between the $T_{mid}$ obtained as a result of fitting the transit light curve and the calculated $T_{(Epoch)}$ obtained from the following ephemeris:

$$T_{(Epoch)} = 2456458.59219(9) + 1.482247(063) \cdot E,$$

where $T_0$ was taken from (Mislis et al., 2015) and *E* is the cycle number.

The measurements of mid-transit moments $T_{mid}$, ratio $R_b/R_*$ and LD *u1* and *u2* coefficients are presented in Table 3. Values of uncertainties were calculated using formulae from (Carter et al., 2008). Also we included in the Table 3 the values of high-precision follow-up photometry of TrES-5b transits from (Mislis et al., 2015 and Maciejewski et al., 2016).

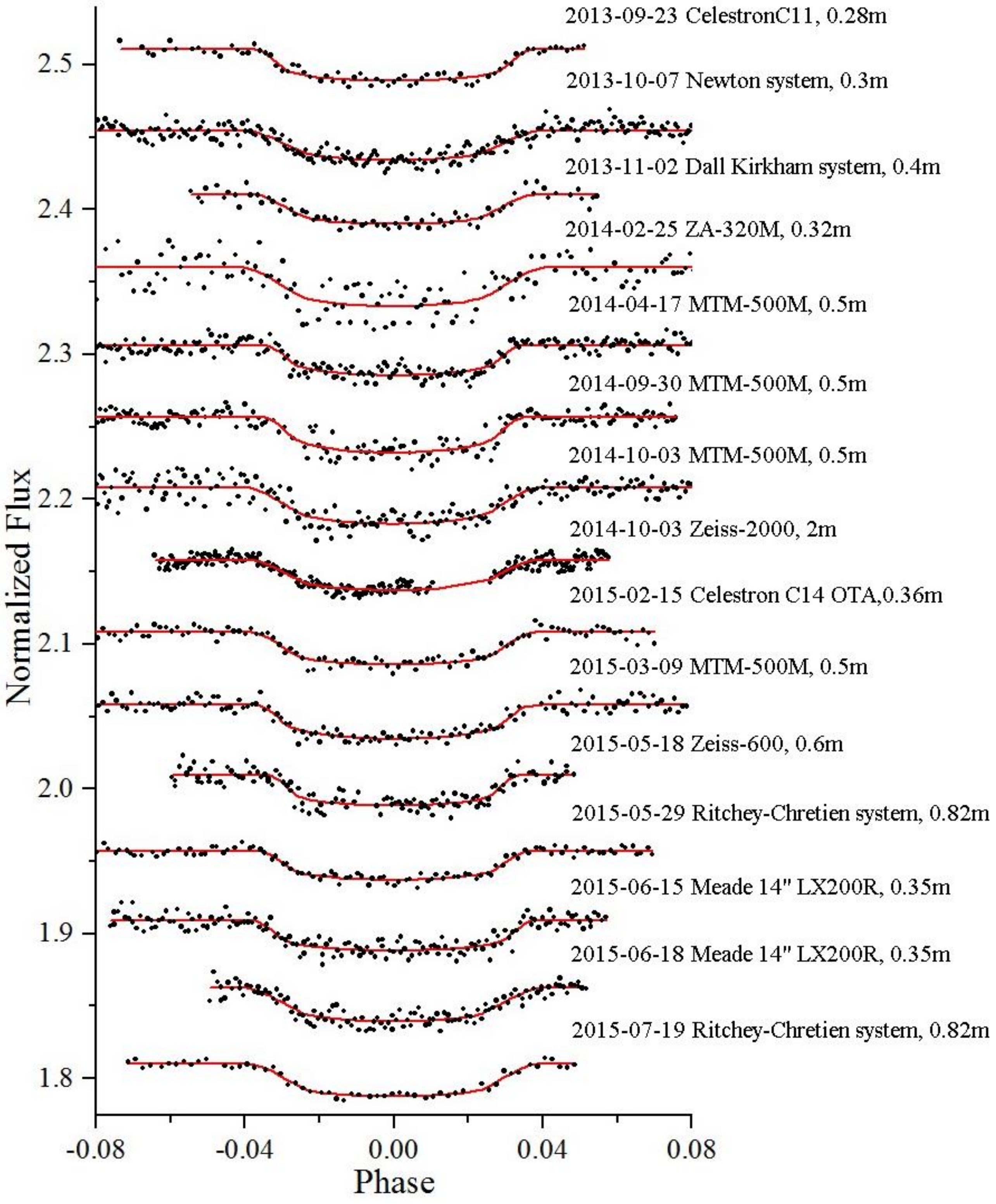

Fig.2(a) Light curves of TrES-5b transits. The best-fit curves are plotted with a red line. Residuals are presented on the bottom panel.

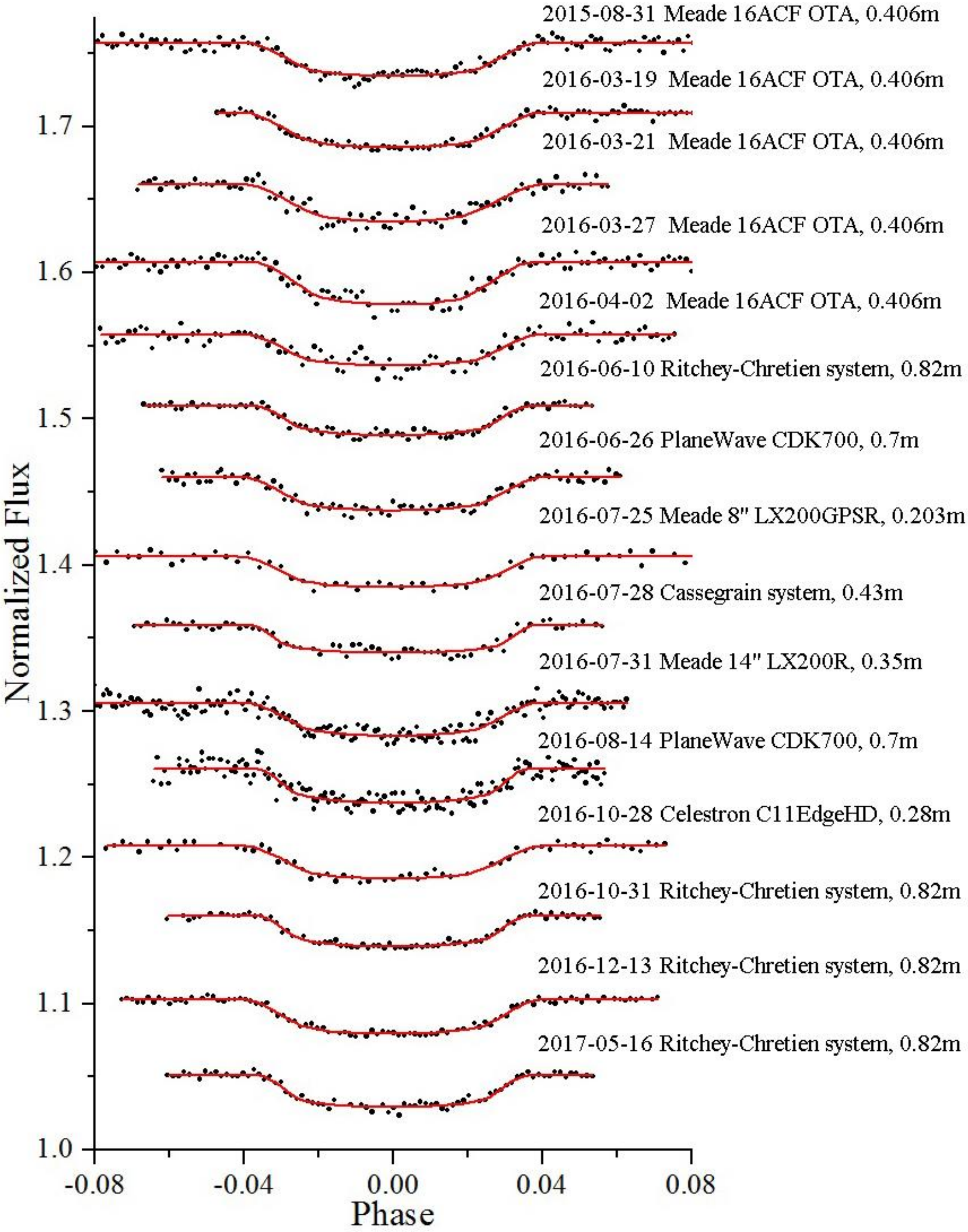

Fig.2(b) Light curves of TrES-5b transits. The best-fit curves are plotted with a red line. Residuals are presented on the bottom panel.

Table 3. The parameter values of the best fitted model of each light curve from this work and works (Mislis et al., 2015 and Maciejewski et al., 2016)

| Date UT | *T*mid (BJD$_{TDB}$2400000+) | *O-C*, d | $R_b/R_*$ | $T_{Dur}$, days | $i_b$ | *u1* | *u2* | Source of data |
|---|---|---|---|---|---|---|---|---|
| 2011.08.26 | $55800.47470^{+0.00023}_{-0.00023}$ | 0.00033 | $0.142^{+0.005}_{-0.005}$ | | $85.0^{+1.4}_{-1.4}$ | - | - | Mislis et al., 2015 |
| 2012.09.10 | $56181.41212^{+0.00029}_{-0.00029}$ | 0.00013 | $0.139^{+0.005}_{-0.005}$ | | $84.9^{+1.0}_{-1.0}$ | - | - | Mislis et al., 2015 |
| 2012.12.08 | $56270.34769^{+0.00080}_{-0.00083}$ | 0.00094 | $0.1327^{+0.0053}_{-0.0039}$ | $0.0754^{+0.0029}_{-0.0022}$ | $86.2^{+2.4}_{-2.4}$ | $0.503^{+0.050}_{-0.049}$ | $0.200^{+0.049}_{-0.049}$ | This work (ETD) |
| 2013.04.09 | $56391.8898^{+0.0016}_{-0.0016}$ | -0.00153 | $0.1453^{+0.012}_{-0.0098}$ | $0.0820^{+0.0088}_{-0.0070}$ | $85.2^{+1.6}_{-2.7}$ | $0.505^{+0.050}_{-0.051}$ | $0.199^{+0.052}_{-0.050}$ | This work (ETD) |
| 2013.06.15 | $56458.59213^{+0.00049}_{-0.00049}$ | -0.00014 | $0.145^{+0.002}_{-0.002}$ | | $84.5^{+0.6}_{-0.6}$ | - | - | Mislis et al., 2015 |
| 2013.07.30 | $56504.54182^{+0.00030}_{-0.00030}$ | -0.00013 | $0.141^{+0.005}_{-0.005}$ | | $85.1^{+0.8}_{-0.8}$ | - | - | Mislis et al., 2015 |
| 2013.05.31 | $56443.7710^{+0.0016}_{-0.0015}$ | 0.00146 | $0.154^{+0.022}_{-0.024}$ | $0.0768^{+0.0064}_{-0.0070}$ | $84.4^{+1.9}_{-3.0}$ | $0.509^{+0.050}_{-0.051}$ | $0.199^{+0.049}_{-0.051}$ | This work (ETD) |
| 2013.08.05 | $56510.4718^{+0.0015}_{-0.0015}$ | 0.00074 | $0.1448^{+0.011}_{-0.00075}$ | $0.0752^{+0.0069}_{-0.0041}$ | $85.1^{+2.2}_{-2.7}$ | $0.507^{+0.050}_{-0.051}$ | $0.200^{+0.050}_{-0.050}$ | This work (ETD) |
| 2013.09.05 | $56541.59774^{+0.00081}_{-0.00093}$ | -0.00023 | $0.1334^{+0.0048}_{-0.0036}$ | $0.0716^{+0.0028}_{-0.0023}$ | $86.2^{+2.5}_{-2.6}$ | $0.493^{+0.050}_{-0.050}$ | $0.196^{+0.051}_{-0.052}$ | This work (ETD) |
| 2013.09.14 | $56550.49157^{+0.00020}_{-0.00021}$ | -0.00005 | $0.184^{+0.0014}_{-0.0014}$ | - | $85.5^{+0.91}_{-0.91}$ | - | - | Mislis et al., 2015 |
| 2013.09.23 | $56596.38431^{+0.00065}_{-0.00061}$ | -0.00081 | $0.1336^{+0.0040}_{-0.0034}$ | $0.0751^{+0.0021}_{-0.0017}$ | $87.1^{+2.0}_{-2.3}$ | $0.500^{+0.049}_{-0.051}$ | $0.200^{+0.049}_{-0.051}$ | This work |
| 2013.10.07 | $56572.7236^{+0.0011}_{-0.0016}$ | -0.00144 | $0.1351^{+0.0077}_{-0.0067}$ | $0.0746^{+0.0056}_{-0.0040}$ | $84.9^{+2.0}_{-3.3}$ | $0.515^{+0.050}_{-0.053}$ | $0.197^{+0.055}_{-0.048}$ | This work |
| 2013.10.30 | $56596.4398^{+0.0011}_{-0.0011}$ | -0.00147 | $0.1283^{+0.0067}_{-0.0060}$ | $0.0715^{+0.0032}_{-0.0028}$ | $86.1^{+2.5}_{-2.5}$ | $0.496^{+0.050}_{-0.051}$ | $0.193^{+0.050}_{-0.050}$ | This work (ETD) |
| 2013.11.02 | $56599.4056^{+0.0011}_{-0.0011}$ | -0.00034 | $0.1305^{+0.0068}_{-0.0062}$ | $0.0718^{+0.0034}_{-0.0028}$ | $85.4^{+2.6}_{-2.7}$ | $0.506^{+0.050}_{-0.051}$ | $0.202^{+0.049}_{-0.049}$ | This work |
| 2013.11.08 | $56605.33486^{+0.00023}_{-0.00021}$ | 0.000221 | - | - | - | - | - | Maciejewski et al., 2016 |
| 2013.12.03 | $56630.5342^{+0.0017}_{-0.0014}$ | 0.00149 | $0.1399^{+0.0092}_{-0.0069}$ | $0.0749^{+0.0053}_{-0.0044}$ | $85.1^{+2.3}_{-2.8}$ | $0.504^{+0.050}_{-0.050}$ | $0.200^{+0.051}_{-0.050}$ | This work (ETD) |

| | | | | | | | | | |
|---|---|---|---|---|---|---|---|---|---|
| 2014.02.25 | $56713.5388^{+0.0021}_{-0.0020}$ | -0.00002 | $0.1497^{+0.012}_{-0.0084}$ | $0.0765^{+0.0077}_{-0.0064}$ | $85.7^{+2.4}_{-2.7}$ | $0.499^{+0.051}_{-0.051}$ | $0.197^{+0.052}_{-0.052}$ | This work |
| 2014.04.17 | $56765.41798^{+0.00065}_{-0.00058}$ | 0.00056 | $0.1275^{+0.0032}_{-0.0029}$ | $0.0685^{+0.0017}_{-0.0016}$ | $87.5^{+1.7}_{-2.4}$ | $0.588^{+0.050}_{-0.052}$ | $0.130^{+0.052}_{-0.050}$ | This work |
| 2014.07.14 | $56852.8695^{+0.0012}_{-0.0012}$ | -0.00056 | $0.1465^{+0.0075}_{-0.0053}$ | $0.0773^{+0.0038}_{-0.0030}$ | $85.8^{+2.3}_{-2.4}$ | $0.506^{+0.050}_{-0.053}$ | $0.201^{+0.050}_{-0.050}$ | This work (ETD) |
| 2014.07.31 | $56870.6547^{+0.0011}_{-0.0012}$ | -0.00243 | $0.1387^{+0.0072}_{-0.0053}$ | $0.0781^{+0.0050}_{-0.0036}$ | $86.0^{+2.2}_{-2.4}$ | $0.500^{+0.053}_{-0.051}$ | $0.196^{+0.050}_{-0.053}$ | This work (ETD) |
| 2014.08.21 | $56891.40861^{+0.00044}_{-0.00043}$ | 0.00005 | - | - | - | - | - | Maciejewski et al., 2016 |
| 2014.08.15 | $56885.47872^{+0.00078}_{-0.00077}$ | -0.00081 | $0.1369^{+0.0043}_{-0.0036}$ | $0.0730^{+0.0022}_{-0.0019}$ | $86.5^{+2.2}_{-2.4}$ | $0.503^{+0.050}_{-0.050}$ | $0.199^{+0.050}_{-0.050}$ | This work (ETD) |
| 2014.09.03 | $56904.74934^{+0.00089}_{-0.00093}$ | 0.00065 | $0.1363^{+0.0057}_{-0.0045}$ | $0.0744^{+0.0030}_{-0.0024}$ | $86.2^{+2.4}_{-2.4}$ | $0.503^{+0.050}_{-0.051}$ | $0.198^{+0.050}_{-0.049}$ | This work (ETD) |
| 2014.09.12 | $56913.64241^{+0.00064}_{-0.00064}$ | 0.00022 | $0.1340^{+0.0045}_{-0.0041}$ | $0.0780^{+0.0025}_{-0.0019}$ | $85.0^{+2.1}_{-2.0}$ | $0.503^{+0.050}_{-0.052}$ | $0.199^{+0.050}_{-0.049}$ | This work (ETD) |
| 2014.09.30 | $56931.43084^{+0.00065}_{-0.00068}$ | 0.00161 | $0.1386^{+0.0039}_{-0.0032}$ | $0.0702^{+0.0025}_{-0.0022}$ | $87.1^{+2.0}_{-2.5}$ | $0.617^{+0.051}_{-0.049}$ | $0.141^{+0.051}_{-0.051}$ | This work |
| 2014.10.03 | $56934.39484^{+0.00063}_{-0.00062}$ | 0.00118 | $0.1430^{+0.0073}_{-0.0056}$ | $0.0749^{+0.0045}_{-0.0034}$ | $85.5^{+2.5}_{-2.5}$ | $0.618^{+0.050}_{-0.051}$ | $0.143^{+0.051}_{-0.051}$ | This work |
| 2014.10.03 | $56934.3945^{+0.00040}_{-0.00038}$ | 0.00088 | $0.1371^{+0.0039}_{-0.0039}$ | $0.0751^{+0.0022}_{-0.0022}$ | $84.1^{+1.5}_{-2.6}$ | $0.496^{+0.049}_{-0.051}$ | $0.199^{+0.05}_{-0.05}$ | This work |
| 2015.02.15 | $57069.27778^{+0.00080}_{-0.00085}$ | -0.00031 | $0.1362^{+0.0050}_{-0.0040}$ | $0.0734^{+0.0027}_{-0.0023}$ | $85.8^{+2.5}_{-2.5}$ | $0.495^{+0.050}_{-0.049}$ | $0.197^{+0.049}_{-0.051}$ | This work |
| 2015.03.09 | $57091.51130^{+0.00067}_{-0.00070}$ | -0.00056 | $0.1376^{+0.0042}_{-0.0034}$ | $0.0752^{+0.0024}_{-0.0020}$ | $86.8^{+2.1}_{-2.2}$ | $0.616^{+0.050}_{-0.050}$ | $0.141^{+0.051}_{-0.050}$ | This work |
| 2015.05.18 | $57161.17656^{+0.00064}_{-0.00067}$ | -0.00095 | $0.1319^{+0.0040}_{-0.0037}$ | $0.0680^{+0.0023}_{-0.0020}$ | $86.7^{+2.2}_{-2.5}$ | $0.478^{+0.050}_{-0.051}$ | $0.189^{+0.049}_{-0.051}$ | This work |
| 2015.05.29 | $57171.55368^{+0.00064}_{-0.00063}$ | 0.00041 | $0.1278^{+0.0042}_{-0.0032}$ | $0.0724^{+0.0022}_{-0.0018}$ | $86.0^{+2.4}_{-2.2}$ | $0.617^{+0.050}_{-0.051}$ | $0.139^{+0.048}_{-0.050}$ | This work |
| 2015.06.15 | $57189.33902^{+0.00070}_{-0.00078}$ | -0.00121 | $0.1306^{+0.0038}_{-0.0033}$ | $0.0762^{+0.0022}_{-0.0019}$ | $87.3^{+1.8}_{-2.3}$ | $0.488^{+0.051}_{-0.049}$ | $0.196^{+0.048}_{-0.050}$ | This work |
| 2015.06.18 | $57192.3031^{+0.0011}_{-0.0016}$ | -0.00136 | $0.1415^{+0.0072}_{-0.0070}$ | $0.0752^{+0.0051}_{-0.0051}$ | $84.5^{+2.0}_{-2.9}$ | $0.500^{+0.051}_{-0.049}$ | $0.196^{+0.051}_{-0.049}$ | This work |
| 2015.07.11 | $57214.54060^{+0.0016}_{-0.00076}$ | 0.00167 | $0.1448^{+0.0030}_{-0.0037}$ | $0.0742^{+0.0021}_{-0.0025}$ | $84.0^{+1.8}_{-3.0}$ | $0.480^{+0.049}_{-0.046}$ | $0.190^{+0.051}_{-0.050}$ | This work (ETD) |
| 2015.07.14 | $57217.5035^{+0.0029}_{-0.0023}$ | 0.00005 | $0.1678^{+0.0055}_{-0.0049}$ | $0.08552^{+0.012}_{-0.011}$ | $83.6^{+2.2}_{-5.1}$ | $0.497^{+0.055}_{-0.053}$ | $0.198^{+0.052}_{-0.052}$ | This work (ETD) |
| 2015.07.19 | $57223.43238^{+0.00086}_{-0.00087}$ | 0.00043 | $0.1368^{+0.0069}_{-0.0048}$ | $0.0733^{+0.0041}_{-0.0028}$ | $85.1^{+2.3}_{-2.4}$ | $0.506^{+0.050}_{-0.051}$ | $0.203^{+0.049}_{-0.049}$ | This work |
| 2015.08.14 | $57248.62942^{+0.00070}_{-0.00077}$ | -0.00074 | $0.1356^{+0.0045}_{-0.0037}$ | $0.0705^{+0.0022}_{-0.0019}$ | $87.0^{+2.1}_{-2.6}$ | $0.496^{+0.049}_{-0.050}$ | $0.193^{+0.052}_{-0.048}$ | This work (ETD) |
| 2015.08.25 | $57260.48794^{+0.00034}_{-0.00032}$ | -0.00026 | - | - | - | - | - | Maciejewski et al., 2016 |
| 2015.08.31 | $57266.41639^{+0.00081}_{-0.0013}$ | -0.00061 | $0.1361^{+0.0075}_{-0.0042}$ | $0.0723^{+0.0037}_{-0.0025}$ | $85.3^{+2.7}_{-2.8}$ | $0.507^{+0.050}_{-0.050}$ | $0.207^{+0.050}_{-0.053}$ | This work |
| 2016.03.19 | $57466.51988^{+0.00038}_{-0.00043}$ | -0.00036 | $0.1410^{+0.0075}_{-0.0055}$ | $0.0750^{+0.0033}_{-0.0027}$ | $84.9^{+2.1}_{-2.4}$ | $0.527^{+0.048}_{-0.052}$ | $0.208^{+0.051}_{-0.050}$ | This work |
| 2016.03.21 | $57469.48333^{+0.00070}_{-0.00073}$ | -0.00139 | $0.1547^{+0.0077}_{-0.0012}$ | $0.0770^{+0.0040}_{-0.0044}$ | $84.5^{+1.8}_{-2.7}$ | $0.507^{+0.052}_{-0.053}$ | $0.200^{+0.048}_{-0.048}$ | This work |
| 2016.03.27 | $57475.41258^{+0.00066}_{-0.00064}$ | -0.00114 | $0.1593^{+0.011}_{-0.0086}$ | $0.0718^{+0.0042}_{-0.0031}$ | $83.8^{+2.2}_{-2.9}$ | $0.511^{+0.051}_{-0.051}$ | $0.201^{+0.052}_{-0.051}$ | This work |
| 2016.04.02 | $57481.34368^{+0.00086}_{-0.00089}$ | 0.00097 | $0.1337^{+0.010}_{-0.0064}$ | $0.0742^{+0.0055}_{-0.0039}$ | $85.2^{+2.1}_{-2.6}$ | $0.508^{+0.051}_{-0.052}$ | $0.201^{+0.051}_{-0.050}$ | This work |
| 2016.06.10 | $57549.52692^{+0.00036}_{-0.0004}$ | 0.00084 | $0.1313^{+0.0048}_{-0.0041}$ | $0.0717^{+0.0025}_{-0.0021}$ | $84.9^{+2.4}_{-2.3}$ | $0.496^{+0.049}_{-0.050}$ | $0.200^{+0.050}_{-0.049}$ | This work |
| 2016.06.26 | $57565.82921^{+0.00057}_{-0.00053}$ | -0.00158 | $0.1427^{+0.0045}_{-0.0055}$ | $0.0774^{+0.0030}_{-0.0031}$ | $84.7^{+1.7}_{-2.8}$ | $0.488^{+0.052}_{-0.048}$ | $0.191^{+0.052}_{-0.051}$ | This work |
| 2016.07.25 | $57595.47545^{+0.00086}_{-0.00082}$ | -0.00029 | $0.1346^{+0.0074}_{-0.0057}$ | $0.0767^{+0.0047}_{-0.0041}$ | $85.4^{+2.0}_{-2.5}$ | $0.498^{+0.049}_{-0.051}$ | $0.197^{+0.049}_{-0.051}$ | This work |
| 2016-07-28 | $57598.43961^{+0.00057}_{-0.00054}$ | -0.00063 | $0.1247^{+0.0037}_{-0.0033}$ | $0.0758^{+0.0023}_{-0.0020}$ | $87.1^{+1.9}_{-2.3}$ | $0.484^{+0.050}_{-0.047}$ | $0.195^{+0.050}_{-0.052}$ | This work |
| 2016-07-31 | $57601.40431^{+0.00059}_{-0.00067}$ | -0.00042 | $0.1391^{+0.0061}_{-0.0060}$ | $0.0740^{+0.0035}_{-0.0032}$ | $84.8^{+2.2}_{-2.6}$ | $0.504^{+0.050}_{-0.051}$ | $0.199^{+0.051}_{-0.050}$ | This work |
| 2016-08-14 | $57614.74652^{+0.00057}_{-0.00056}$ | 0.00156 | $0.1388^{+0.0069}_{-0.0042}$ | $0.0726^{+0.0025}_{-0.0019}$ | $86.2^{+2.4}_{-2.6}$ | $0.509^{+0.052}_{-0.051}$ | $0.201^{+0.051}_{-0.049}$ | This work |
| 2016-10-28 | $57690.33979^{+0.00060}_{-0.00061}$ | 0.00024 | $0.1389^{+0.0071}_{-0.0065}$ | $0.0746^{+0.0044}_{-0.0031}$ | $84.9^{+2.1}_{-2.4}$ | $0.504^{+0.052}_{-0.053}$ | $0.199^{+0.050}_{-0.048}$ | This work |
| 2016-10-31 | $57693.30435^{+0.00025}_{-0.00030}$ | 0.00030 | $0.1305^{+0.0039}_{-0.0025}$ | $0.0711^{+0.0016}_{-0.0014}$ | $85.8^{+2.4}_{-2.0}$ | $0.524^{+0.048}_{-0.048}$ | $0.204^{+0.049}_{-0.051}$ | This work |
| 2016-12-13 | $57736.28968^{+0.00027}_{-0.00026}$ | 0.00046 | $0.1428^{+0.0040}_{-0.0052}$ | $0.0777^{+0.0020}_{-0.0023}$ | $84.7^{+1.6}_{-2.7}$ | $0.544^{+0.050}_{-0.051}$ | $0.222^{+0.049}_{-0.051}$ | This work |
| 2017-05-16 | $57890.44423^{+0.00033}_{-0.00039}$ | 0.00130 | $0.1376^{+0.0039}_{-0.0033}$ | $0.0713^{+0.0020}_{-0.0018}$ | $85.0^{+2.4}_{-2.0}$ | $0.413^{+0.048}_{-0.048}$ | $0.232^{+0.048}_{-0.051}$ | This work |
| Weighted average values this work | | | $0.1405 \pm 0.0011$ | $0.0759 \pm 0.0022$ | $85.78 \pm 0.39$ | | | |
| from literature | | | $0.1475 \pm 0.0009$ | | $84.03 \pm 0.16$ | | | |

## 4 Simulation of a three-body system (star-planet-planet)

We carried out a frequency analysis for transit timing data sets including 45 measurements of *O*-*C* (Observation – Calculation) obtained from the light curves in this work and 8 measurements from (Mislis et al., 2015 and Maciejewski et al., 2016) (53 values in total) with the average σ = 1.1 min. For the analysis we took into account the weights of the measurements and used the "Clean" method, suggested in 1974 by Hogbomom for the cleaning "dirty maps" that are obtained during aperture synthesis in radio astronomy (Hogbom, 1974). Subsequently the method was modified to obtain "clean" spectra in the spectral analysis of time series (Roberts et al., 1987).

The frequency analysis detected a peak at P ~ 99 days. The false alarm probability (FAP) is about 0.18%. After "cleaning" the spectrum by means of the algorithm of the "Clean" method, no evidence was found for equivalent or greater importance peaks. The periodogram is shown in Fig. 3.

The detected peak at P ~ 99 days gives us reason to assume that there is an additional body in the TrES-5 planetary system. To search for it and estimate its mass, as well as the distance from the planet TrES-5b, it was necessary to conduct a dynamic simulation of a possible system consisting of three bodies.

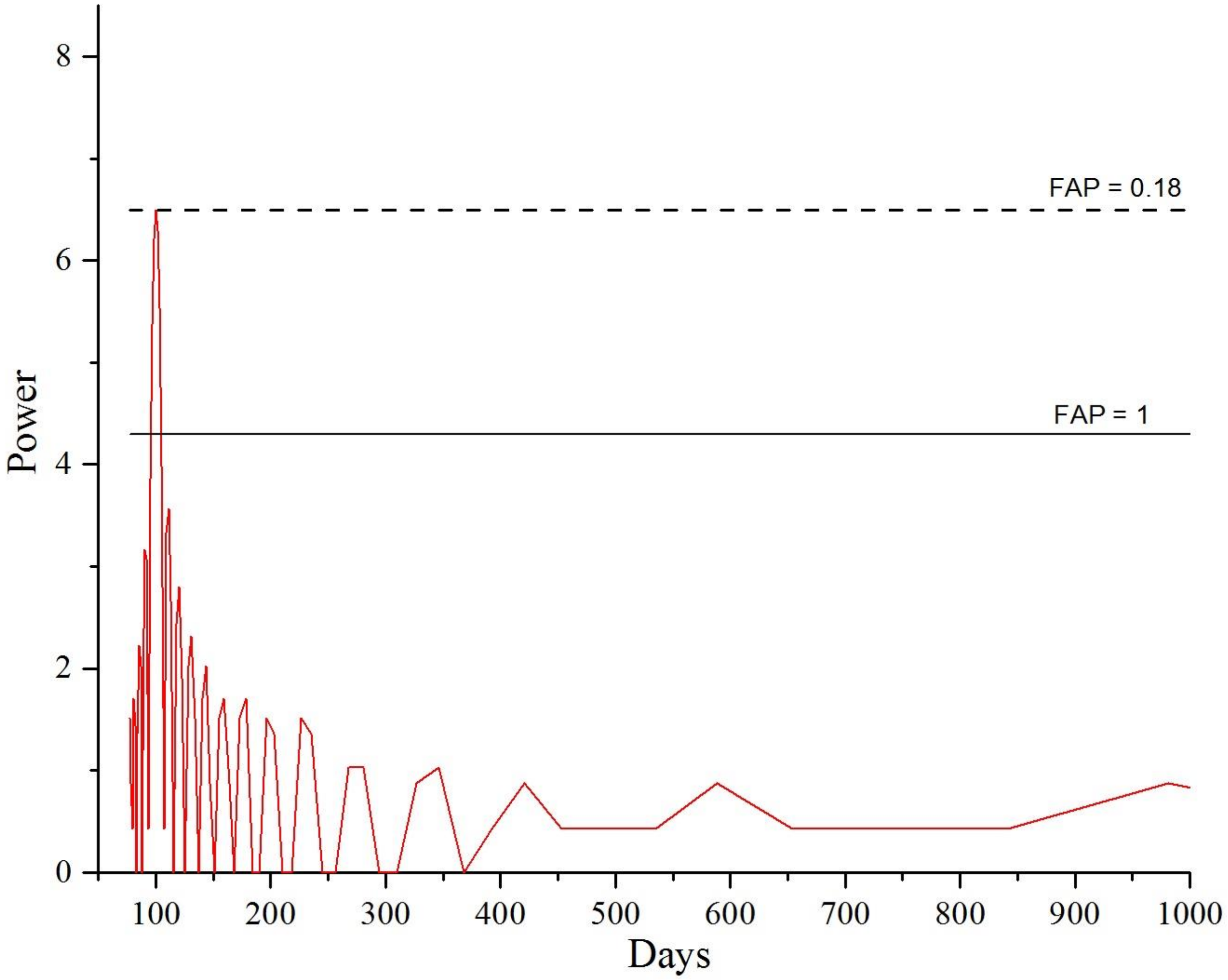

Fig 3. Periodogram of the clean spectrum for the *O-C* data with a peak at a value of 99 days. Dashed line shows probability with FAP=1%. Solid line shows probability of the detected peak with FAP=0.18%.

To construct a dynamic model for a triple system "star-planet-planet", we used translational and rotational motion equations for the two and three body problem obtained by G.N. Duboshin (Duboshin, 1963).

We used a model in which the motion of three bodies in space is simulated. The shape of such bodies cannot be considered as material points, because the force of interaction between them essentially depends on their relative orientation. Thus, their prograde and retrograde motion must be considered together.

This problem of prograde-retrograde motion was and continues to be developed in different assumptions about the parameters of the considered systems. In this numerical investigation of motion in a binary or triple system, each body is considered as a homogeneous triaxial ellipsoid. Differential equations of motion for this system were obtained by G.N. Duboshin (Duboshin, 1963). They are derived from the general second-order Lagrange equations $\frac{d}{dt}\left(\frac{\partial T}{\partial q_i'}\right) - \frac{\partial T}{\partial q_i} = \frac{\partial U}{\partial q_i}$, where for the generalized coordinates $q_i$ we accepted the absolute rectangular coordinates of the inertia centers $(x_i, y_i, z_i)$, describing the prograde and retrograde motion, and the Euler angles $(\varphi_i, \psi_i, \theta_i)$ describing the rotation of the body.

In this investigation, the three-body problem was considered for the simulation of a system with a star in the center and two planets orbiting it. The problem was solved in relative coordinates, with the origin placed in the center of the star. Thus, for this problem, the final form of the above equations is as follows:

$$x_i' = V_{x_i}$$

$$y_i' = V_{y_i}$$

$$z_i' = V_{z_i}$$

$$V_{x_i}' = \frac{(m_0 + m_i)}{m_0 m_i}\frac{\partial U_{i0}}{\partial x_i} + \frac{\partial R_i}{\partial x_i}$$

$$V_{y_i}' = \frac{(m_0 + m_i)}{m_0 m_i}\frac{\partial U_{i0}}{\partial y_i} + \frac{\partial R_i}{\partial y_i}$$

$$V_{z_i}' = \frac{(m_0 + m_i)}{m_0 m_i}\frac{\partial U_{i0}}{\partial z_i} + \frac{\partial R_i}{\partial z_i}$$

(1)

$$A_i p_i' - (B_i - C_i) q_i r_i = (\frac{\partial U}{\partial \psi_i} - \cos\theta_i \frac{\partial U}{\partial \varphi_i})\frac{\sin\varphi_i}{\sin\theta_i} + \cos\varphi_i \frac{\partial U}{\partial \theta_i}$$

$$B_i q_i' - (C_i - A_i) r_i p_i = (\frac{\partial U}{\partial \psi_i} - \cos\theta_i \frac{\partial U}{\partial \varphi_i})\frac{\cos\varphi_i}{\sin\theta_i} - \sin\varphi_i \frac{\partial U}{\partial \theta_i}$$

$$C_i r_i' - (A_i - B_i) p_i q_i = \frac{\partial U}{\partial \varphi_i}$$

$$p_i = \psi_i' \sin\varphi_i \sin\theta_i + \theta_i' \cos\varphi_i$$

$$q_i = \psi_i' \cos\varphi_i \sin\theta_i - \theta_i' \sin\varphi_i$$

$$r_i = \psi_i' \cos\theta_i + \varphi_i'$$

$$(i = 0, 1, 2)$$

The following designations are used: $m_i$ — the mass of the corresponding body, $A_i, B_i, C_i$ — the main central moments of inertia, $p_i, q_i, r_i$ — the projected angular rotation velocity of a body in its own coordinate system, related to the Euler angles through the kinematic equations [Duboshin 1963] ], $R_i$ — perturbation function, which is calculated from the potential $U_{ij}$ :

$$R_i = \sum_{j=1}^{n'} \left( \frac{1}{m_i} U_{ij} + \frac{1}{m_0} \left( x_i \frac{\partial U_{j0}}{\partial x_j} + y_i \frac{\partial U_{j0}}{\partial y_j} + z_i \frac{\partial U_{j0}}{\partial z_j} \right) \right) \quad (2)$$

To calculate the potential, we took the members up to the third order inclusive in the decomposition proposed by G.N. Duboshin:

$$U_{ij} \cong Gm_i m_j + Gm_i \frac{A_j + B_j + C_j - 3I_j^{ij}}{2\Delta_{ij}^3} + Gm_j \frac{A_i + B_i + C_i - 3I_i^{ij}}{2\Delta_{ij}^3} , \quad (3)$$

where $G$ is the gravitational constant, $\Delta_{ij} = \sqrt{(x_i - x_j)^2 + (y_i - y_j)^2 + (z_i - z_j)^2}$ is the distance between the centers of the bodies, and $I_s^{ij}$ is the moment of inertia relative to the line connecting the centers of inertia of the two bodies. It should be noted that this approximation of the potential works well provided that the distance between the bodies is larger than their size. For the objects under investigation, this condition is generally met.

The system of equations (1) is a system of differential equations of the 1st order. To obtain its numerical solution, the Dormand - Prince integration method was used, which is based on the 8th order Runge - Kutta method (Haireret al, 1993). The integration accuracy was ~$10^{-7}$ km. The criterion of a successful implementation of the numerical integration was the constancy of the classical integrals of the system (1) - areas and energy. The accuracy of the results was determined by integrating in the forward and reverse directions. At the same time, the parameters obtained as a result of the reverse integration were compared with the initial conditions.

For the initial simulation parameters, we used the mass of the TrES-5b, the mass of the parent star $M_* = 0.893$ (± 0.024) $M_{Sol}$ obtained in (Mandushev et. al., 2011), and the re-determined value of $P_b$.

The mass of the 3-rd body in the system was set in the range of the mass of the Mars M ≈ $0.1M_{Earth}$ to the mass of brown dwarf M = 30 $M_{Jup}$. The simulation was performed at the resonances 1:2, 2:3, 1:3, 3:4, 2:5, 3:5 and 4:5. Thus, we iteratively selected model parameters that would provide the best agreement with the observational data presented in the *O-C* diagram. The resulting model-based timing at the resonances 2:3, 3:4, 2:5, 3:5 and 4:5, with the period P ~ 99 days, was obtained with an amplitude much greater than expected. A further increase of the semi-major axis of the 3-rd body, i.e. at the potential resonances 1:4, 1:5, 1:6, etc., in the system would give us a progressive increase of mass estimates for the 3-rd object reaching to the mass of a brown dwarf. Wherein the presence of a third body with a mass comparable to the mass of a brown dwarf in an orbit close to TrES-5b's orbit would be easy to register with the only 8 currently available radial velocity measurements presented in Mandushev et. al., (2011).

Based on all the considered resonances with masses in the range of $0.1M_{Earth}$ to 30 $M_{Jup}$, the best agreement of model and observed data was obtained for 2 cases:

- Resonance 1:2 with the mass of the third body $M_{Planet_2} \sim 0.24M_{Jup}$;
- Resonance 1:3 with the mass of the third body $M_{Planet_2} \sim 3.15M_{Jup}$.

The case with $M_{Planet_2} \sim 3.15M_{Jup}$ cannot be considered further because of the limitations of the radial velocities registered by Mandushev et. al. (2011) for TrES-5b. An object of such mass orbiting around the star with a 1:3 resonance would produce radial velocities exceeding 400 m/s, that could be simply detected based on the RV analysis.

As the result, Fig. 4 shows the simulated transit timing of TrES-5b interacting with a 3-rd body in the system. For the model and all presented in the Table 3 observations the reduced $\chi^2_{Model}$ = 0.32, whereas for the case of linear ephemerides $\chi^2_{Lin}$ = 0.57. Thus, it can be argued that our model curve (red series - Fig. 4) based upon a 1:2 resonance and ~99-day period agrees better with the distribution of observations (points - Fig. 4) than the linear model.

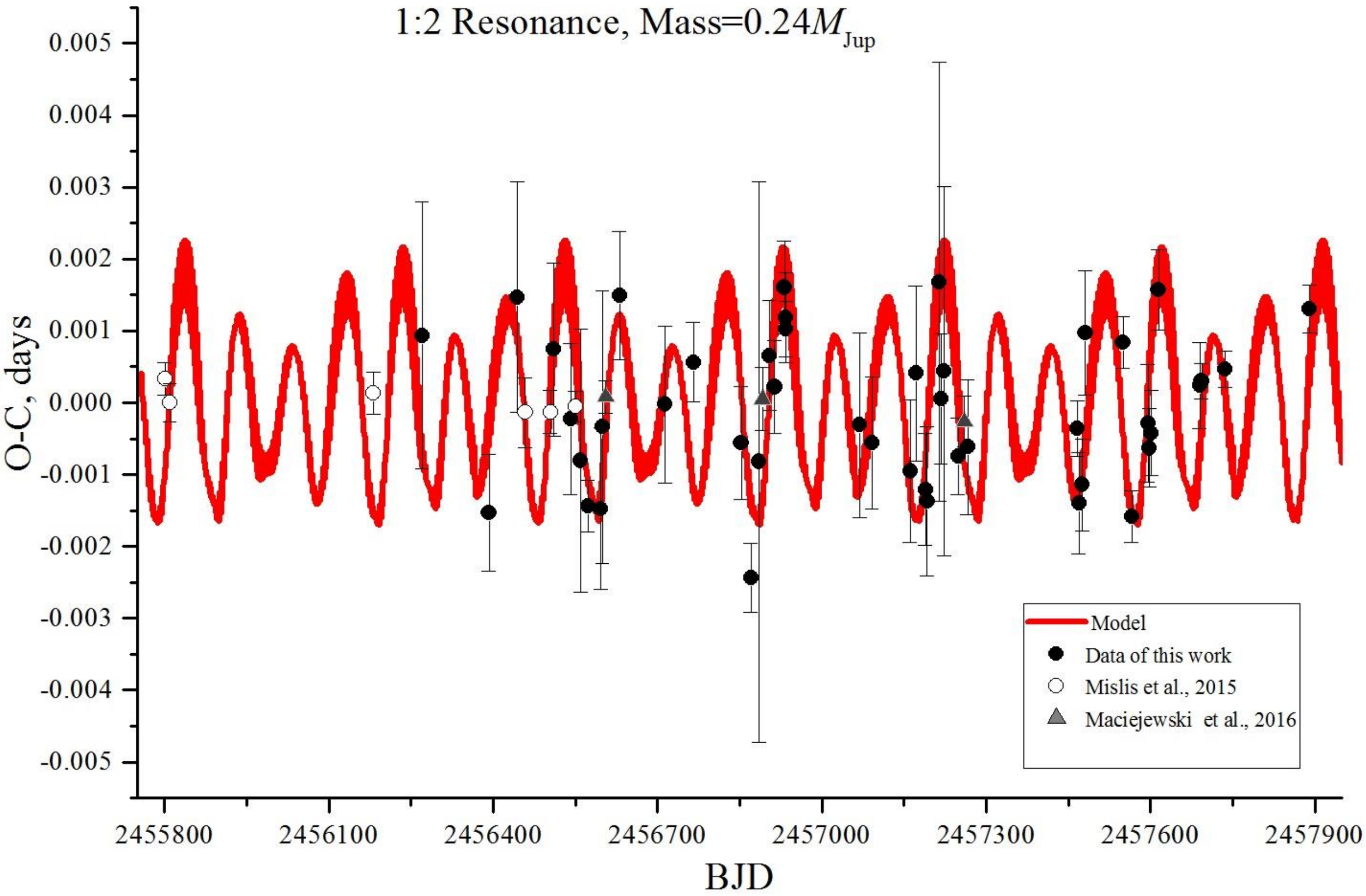

Fig. 4. Observed data with a superimposed model curve. Black points – the observations from this work; white points – data from (Mislis et al., 2015); grey triangles – data from (Maciejewski et al., 2016).

## 6 Radial velocities analysis with data from literature

For the radial velocities (RV) analysis of the host star of TrES-5b we searched data in the literature and RV archives. There are only 8 measurements of RV of the star TrES-5 presented in (Mandushev et. al., 2011).

We analyzed available set of RV data using the Markov Chain Monte-Carlo (MCMC) code described in (Gillon et al. 2012). This software uses a Keplerian model of (Murray & Correia, 2010) to fit the RVs. We obtained the physical parameters of the planet from the set of the parameters that were perturbed randomly at each step of the Markov chains (jump parameters), stellar mass and radius. Free eccentricity was assumed. The prior physical parameters of the star log $g$ = 4.517 ± 0.012, $T_{eff}$ = 5171 ± 36, [Fe/H] = 0.2 ± 0.1 were used. As for the orbital period of TrES5b modelling, we used fixed value $P_b$ = 1.482247063 days.

As the result of the modeling, we obtained the best fit-model with $\chi^2$ = 4.5 for the eccentricity $e$ = 0.017 ± 0.012. The planetary parameters are presented in Table 4.

Table 4. The planetary parameters for the model with fixed re-determined period $P_b$

| **Parameter** | **Value** | **Units** |
|---|---|---|
| Period, $P_b$ | 1.482247063 ± 0.0000005 | days |
| Eccentricity, $e$ | 0.017 ± 0.012 | |
| semi-major axis, $a$ | 0.02447 ± 0.00021 | AU |
| RV semi-amplitude, $K$ | 343 ± 11 | m/s |
| Minimum mass, $M_P \sin i$ | 1.784 ± 0.066 | $M_{Jup}$ |

The plot of the model with the residuals for the eight RV measurements are presented in Figure 5. The RMS of the fitting procedure is 20 m/s and the maximum deviation from this model reaches 36.3 m/s.

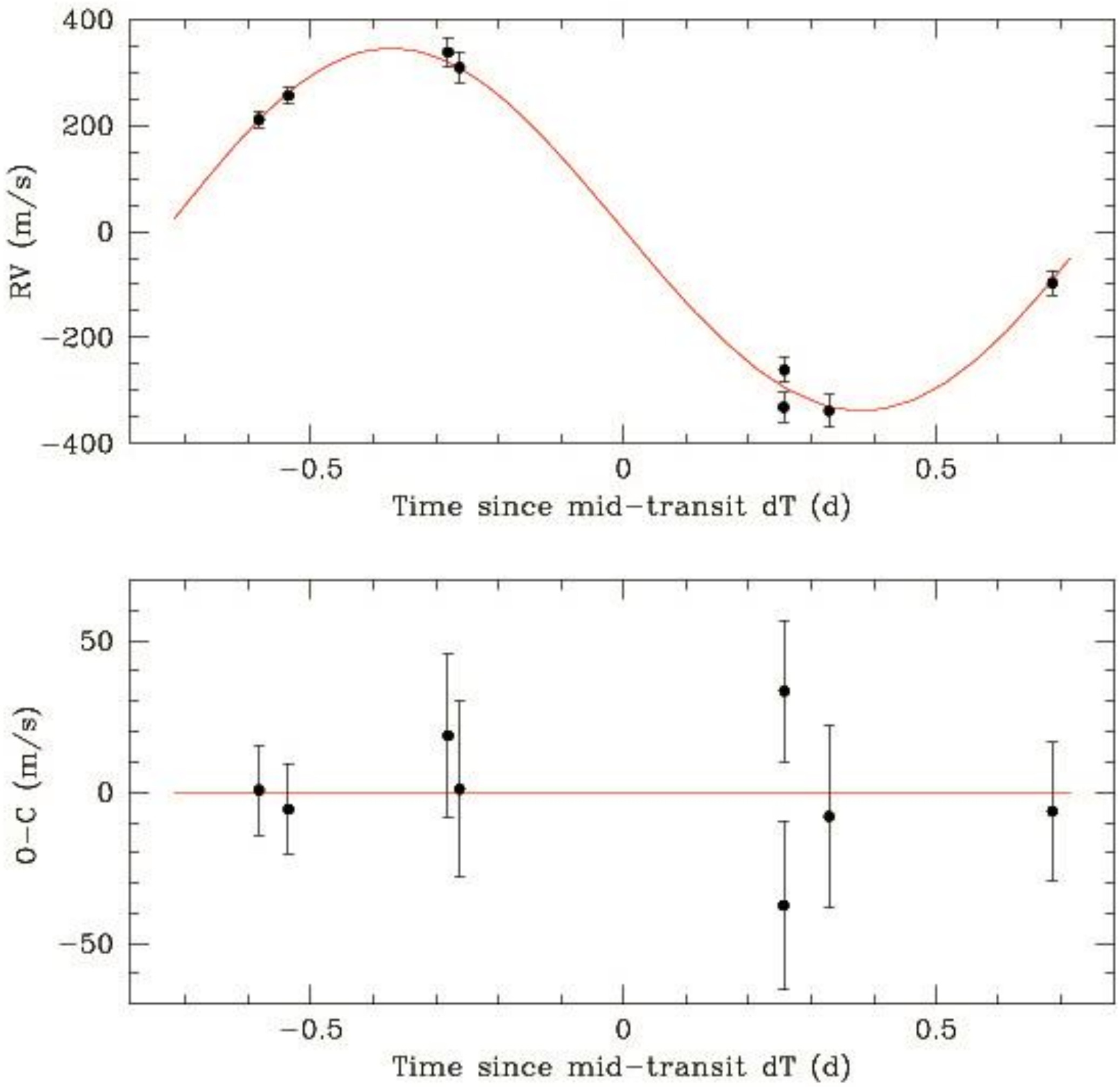

Fig 5. Top: (black points) radial velocities with uncertainties for the star TrES-5 from (Mandushev et. al., 2011) with (solid line) best fit-model to the eight radial velocities for the fixed orbital period of TrES-5b $P_b$ = 1.482247063 days. Bottom: The residuals from the best fit-model and radial velocities.

When carrying out a similar analysis of radial velocities using the period $P_b$ = 1.4822446 days presented in (Mandushev et. al., 2011), the best fit-model is $\chi^2$ = 6. Thus our model with $\chi^2$ = 4.5 gives orbital parameters of TrES-5b that are a little more accurate when compared with the model of (Mandushev et. al., 2011).

## 7 Discussion and conclusions

Based on an analysis of the photometric observations of transits of TrES-5b, obtained as part of EXPANSION project to study the TTV of the exoplanet, with the use data from the ETD and high-precision photometry from (Mislis et al., 2015 and Maciejewski et al., 2016), transit timing variations of TrES-5b with a period of about 99 days was detected.

The resulting speckle-interferometric observations with the 6-meter BTA telescope allow us to confidently announce the absence of any objects close to the host star with a brightness difference of Δm: 0mag ÷ 1mag and in the distance range of ρ: 200 mas ÷ 3000 mas. This fact indicates the absence of any components near TrES-5 of stellar masses greater than the mass of a brown dwarf at distances 72 AU ÷ 1080 AU.

To estimate the mass and calculate the orbital parameters for the third component in the system perturbing the orbit of TrES-5b, we conducted an N-body simulation at the resonances 1:2, 2:3, 1:3, 3:4, 2:5, 3:5 and 4:5.

Based on the conducted N-body simulation we detected the simulated transit timing variations for a perturbing Neptune mass body at the 1:2 resonance are in good agreement with the period P ~ 99 days, amplitude, and profile obtained from the TrES-5b observations. Thus we were able to predict a possible existence of planet TrES-5c with a mass $M_{TrES\text{-}5c} \sim 0.24 M_{Jup}$ at the 1:2 resonance to TrES-5b.

At the same time, on the other resonances, taking into account the correlation between observations and N-body simulation, and also based on the radial velocities analysis of the parent star, we did not find any evidence for the existence of other bodies in the system close to the orbit of Tres-5b.

It should be noted that the estimate of the radial velocity for a planet with a mass of 0.24 $M_{Jup}$ with the orbital period of 2.96 days (which corresponds to a resonance of 1:2) would produce an RV variation with semi-amplitude of about 35-40 m/s for a circular orbit. On the basis of only 8 measurements of the radial velocities of Tres-5 presented in (Mandushev et. al., 2011), we cannot conduct a search for a secondary planet in this system. But the results of our RV analysis of the RMS (20 m/s) and the maximum deviation of the observed values from the model-fit curve (36 m/s) model may indicate the existence of additional perturbations in the system that cannot be explained by the only exoplanet investigated in the system.

To verify the possible existence or absence of the exoplanet TrES-5c, additional high-precision radial velocity and photometric measurements of TrES-5 are necessary.

## Acknowledgments

The photometric and speckle-interferometric observations was supported by the Russian Science Foundation grant No. 14-50-00043. Theoretical investigations were supported by Russian Foundation for Basic Research (project No. 17-02-00542). This paper makes use of EXOFAST (Eastman et al. 2013) as provided by the NASA Exoplanet Archive, which is operated by the California Institute of Technology, under contract with the National Aeronautics and Space Administration under the Exoplanet Exploration Program. In addition, we thank Vladimir Gerasichev, Vladimir Kouprianov and Carl Knight for much help in preparing the article.